# Multi-objective Optimization For The Dynamic Multi-Pickup and Delivery Problem with Time Windows

**I. Harbaoui Dridi**[(1)][(2)], **R. Kammarti**[(2)]

[(1)]LAGIS : Ecole Centrale de Lille, Villeneuve d'Ascq, France
[(2)] LACS : Ecole Nationale des Ingénieurs de Tunis, Tunis - Belvédère. TUNISIE
imenharbaoui@gmail.com, kammarti.ryan@planet.tn

**P. Borne**[(1)], **M. Ksouri**[(2)]

[(1)]LAGIS : Ecole Centrale de Lille, Villeneuve d'Ascq, France
[(2)] LACS : Ecole Nationale des Ingénieurs de Tunis, Tunis - Belvédère. TUNISIE
p.borne@ec-lille.fr, Mekki.Ksouri@insat.rnu.tn

*Abstract*— *The PDPTW is an optimization vehicles routing problem which must meet requests for transport between suppliers and customers satisfying precedence, capacity and time constraints. We present, in this paper, a genetic algorithm for multi-objective optimization of a dynamic multi pickup and delivery problem with time windows (Dynamic m-PDPTW). We propose a brief literature review of the PDPTW, present our approach based on Pareto dominance method and lower bounds, to give a satisfying solution to the Dynamic m-PDPTW minimizing the compromise between total travel cost and total tardiness time.*

## I.  INTRODUCTION

The vehicle routing problems are studied because of the growing passenger and freight today. Some studies are primarily oriented towards solving the vehicle routing problem (VRP: Vehicle Routing Problem).

Another major area of research focused on an important variant of VRP is the PDPTW, where in addition to the existence of time constraints and capacity constraints on the vehicle, this problem involves a clients set and a suppliers set geographically located. Each routing will also satisfy precedence constraints to ensure that a customer should not be visited before his supplier. The PDPTW is the Pickup and Delivery Problem with Time Windows, is divided into two categories: 1-PDPTW (a vehicle) and m-PDPTW (multiple vehicles). These can also be treated in two versions: static or dynamic

In this paper we present a literature review of the PDPTW followed our multi-objective approach, minimizing the compromise between total travel cost and total tardiness time, based on genetic algorithms and lower bounds, which treats the dynamic m-pdptw by incorporating unexpected while routing have been planned and that their operation has started.

## II.  LITERATURE REVIEW

### A.  Vehicle routing problem

The Vehicle Routing Problem (VRP) represents a multi-goal combinatorial optimization problem which has been the subject of many work and variations in the literature. [1][2]

The VRP principle is: given a depot D and a set of customers orders C = (c1, ... , Cn), to build a package routing, for a finite number of vehicles, beginning and ending at a depot. In these routing, a customer must be served only once by a single vehicle and vehicle capacity transport for a routing should not be exceeded. [3]

The Meta heuristics were also applied to solve the vehicle routing problem.  Among these methods, we can include ant colony algorithms and genetic algorithm which were used for the resolution of DVRP. [4] [5]

Savelsbergh and al have shown that the VRP is a NP-hard problem [6]. Since the m-PDPTW is a generalization of the VRP it's a NP-hard problem.

### B.  The PDPTW: Pickup and Delivery Problem with Time Windows

The PDPTW is a variant of VRPTW where in addition to the existence of time constraints, this problem implies a set of customers and a set of suppliers geographically located. Every routing must also satisfy the precedence constraints to ensure that a customer should not be visited before his supplier. [7]

A dynamic approach for resolving the 1-PDP without and with time windows was developed by Psaraftis, H.N considering objective function as a minimization weighting of the total travel time and the non-customer satisfaction. [8]

Jih, W and al have developed an approach based on the hybrid genetic algorithms to solve the 1-PDPTW, aiming to minimize combination of the total cost and total waiting time. [9]

Another genetic algorithm was developed by Velasco, N and al to solve the 1-PDP bi-objective in which the total travel time must be minimized while satisfy in prioritise the most urgent requests. In this literature, the method proposed to resolve this problem is based on a No dominated Sorting Algorithm (NSGA-II). [10]

Kammarti, R and al deal the 1-PDPTW, minimizing the compromise between the total travel distance, total waiting time and total tardiness time, using an evolutionary algorithm with Special genetic operators, tabu search to provide a set of viable solutions. [11][12]

This work has been extended, in proposing a new approach based on the use of lower bounds and Pareto dominance method, to minimize the compromise between the total travel distance and total tardiness time. [13][14]

About the m-PDPTW, Sol, M and al have proposed a branch and price algorithm to solve the m-PDPTW, minimizing the vehicles number required to satisfy all travel demands and the total travel distance. [15]

Quan, L and al have presented a construction heuristic based on the integration principle with the objective function, minimizing the total cost, including the vehicles fixed costs and travel expenses that are proportional to the travel distance. [16]

A new metaheuristic based on a tabu algorithm, was developed by Li, H and al to solve the m-PDPTW. [17]

Li, H and al have developed a "Squeaky wheel" method to solve the m-PDPTW with a local search. [18]

A genetic algorithm was developed by Harbaoui Dridi. I and al treating the m-PDPTW to minimize the total travel distance and the total transport cost [19]. This work has been extended, in proposing a new approach based on the use Pareto dominance method to give a set of satisfying solutions to the m-PDPTW minimizing total travel cost, total tardiness time and the vehicles number. [20] [21]

One more approach based on the genetic algorithm was also developed by [22] to solving the DPDPTW. This approach is evaluated using a large number of problem instances with varying degrees of dynamism.

Mitrovic-Minic S and al describe a double-horizon based heuristics for the dynamic PDPTW. The Computational results of this work show the advantage of using a double-horizon in conjunction with insertion and improvement heuristics.[23]

Rusdiansyah. A and al have developed a model and heuristic algorithm for Dynamic Pick Up and Delivery Problem with Time Windows for City-Courier providers. Considering the number of timeslots and Degree of Dynamism has a direct relationship to the time required computational.[24]

III. MATHEMATICAL FORMULATION

Our problem is characterized by the following parameters:

- $N$ : Set of customers, supplier and depot vertices,
- $N'$ : Set of customers and supplier vertices,
- $N^+$ : Set of supplier vertices,
- $N^-$ : Set of customers vertices,
- $K$ : Vehicle number,
- $d_{ij}$ : Euclidian distance between the vertex $i$ and the vertex $j$. If $d_{ij} = \infty$ then the road between $i$ and $j$ doesn't exist,
- $t_{ijk}$ : Time used by the vehicle $k$ to travel from the vertex $I$ to the vertex $j$,
- $[e_i, l_i]$ : Time window of the vertex $i$,
- $s_i$ : Stopping time at the vertex $i$,
- $q_i$ : Goods quantity of the vertex $i$ request. If $q_i > 0$, the vertex $i$ is a supplier; if $q_i < 0$, the vertex $i$ is a customer and if $q_i = 0$ then the vertex was served.
- $Q_k$ : Capacity of vehicle $k$,
- $i = 0..N$ : Predecessor vertex index,
- $j = 0..N$ : Successor vertex index,
- $k$: 1..K: Vehicle index,
- $Xijk = \begin{cases} 1 & \text{If the vehicle travel from the vertex } i \text{ to the vertex } j \\ 0 & \text{Else} \end{cases}$
- $A_i$ : Arrival time of the vehicle to the vertex $i$,
- $D_i$ : Departure time of the vehicle from the vertex $i$,
- $y_{ik}$ : The goods quantity in the vehicle $k$ visiting the vertex i,
- $C_k$ : Travel cost associated with vehicle k,
- A vertex is served only once,
- There is one depot,
- The capacity constraint must be respected,

- The depot is the start and the finish vertex for the vehicle,
- The vehicle stops at every vertex for a period of time to allow the request processing,
- If the vehicle arrives at a vertex *i* before its time windows beginning date $e_i$, it waits.

The function to minimize is given as follows:

$$Minimize\ f = \begin{pmatrix} \lambda_1 c_1 \sum_{k \in K} \sum_{i \in N} \sum_{j \in N} C_k d_{ijk} X_{ijk} + \\ \lambda_2 c_2 \sum_{k \in K} \sum_{i \in N} \sum_{j \in N} max(0, D_i - l_i) X_{ijk} \end{pmatrix}$$

(1)

Subject to:

$$\sum_{i=1}^{N} \sum_{k=1}^{K} x_{ijk} = 1, j = 2,...N \quad (2)$$

$$\sum_{j=1}^{N} \sum_{k=1}^{K} x_{ijk} = 1, i = 2,...N \quad (3)$$

$$\sum_{i \in N} X_{i0k} = 1, \forall k \in K \quad (4)$$

$$\sum_{j \in N} X_{0jk} = 1, \forall k \in K \quad (5)$$

$$\sum_{i \in N} X_{iuk} - \sum_{j \in N} X_{ujk} = 0, \forall k \in K, \forall u \in N \quad (6)$$

$$X_{ijk} = 1 \Rightarrow y_{jk} = y_{ik} + q_i, \forall i, j \in N; \forall k \in K \quad (7)$$

$$y_{0k} = 0, \forall k \in K \quad (8)$$

$$Q_k \geq y_{ik} \geq 0, \forall i \in N; \forall k \in K \quad (9)$$

$$D_w \leq D_v, \forall i \in N; \forall w \in N_i^+; \forall v \in N_i^- \quad (10)$$

$$D_0 = 0 \quad (11)$$

$$X_{ijk} = 1 \Rightarrow e_i \leq A_i \leq l_i, \forall i, j \in N; \forall k \in K \quad (12)$$

$$X_{ijk} = 1 \Rightarrow e_i \leq A_i + s_i \leq l_i, \forall i, j \in N; \forall k \in K; s_i \neq 0 \quad (13)$$

$$X_{ijk} = 1 \Rightarrow D_i + t_{ijk} \leq (l_j - s_j), \forall i, j \in N; \forall k \in K \quad (14)$$

The constraint (2) and (3) ensure that each vertex is visited only once by a single vehicle. The constraint (4) and (5) ensure that the vehicle route beginning and finishing is the depot. The constraint (6) ensures the routing continuity by a vehicle.

(7), (8) and (9) are the capacity constraints. The precedence constraints are guaranteed by (10) and (11). The constraints (12), (13) and (14) ensure compliance time windows.

Where $\lambda_i$ and $c_i$ are weights and scaling coefficients.

## IV. MULTI-OBJECTIVE APPROACH FOR THE DYNAMIC m-PDPTW

### A. Using genetic algorithm

The principle of different genetic operations such as chromosome coding, the generation of populations as well as procedures for correcting capacity and precedence are detailed in our work [19].

In our case, we will generate two types of populations. A first population noted $P_{node}$, which represents all nodes to visit with all vehicles, according to the permutation list coding. The second population noted $P_{vehicle}$ indicates nodes number visited by each vehicle.

Whereas these two populations and correction procedures, we obtain the final population $P_{node\ /\ vehicle}$ whose shows an individual example in Fig 1.

| $V_1$ | $C_1$ | 0 | 5 | 8 | 2 | 6 | 4 | 3 | 0 |
| $V_2$ | $C_2$ | 0 | 10 | 7 | 9 | 1 | 0 | | |

Figure 1: Example of Individual of the population $P_{node\ /\ vehicle}$

With: $N' = 10$ and $K=2$

### B. Multi-criteria evaluation

A multi-criteria problem is defined as an optimization vector problem, which seeks to optimize several components of a vector function cost.

However, it is necessary to find solutions representing a possible compromise between the criteria. The Pareto optimality concept introduced by the economist V. Pareto in the nineteenth century is frequently used [25].

V. Pareto formulated the following concept: in a multi-criteria problem, there is a balance so that we cannot improve one criterion without deteriorating at least one other. This balance has been called Pareto optimal

A solution is noted Pareto optimal if it is dominated by any other point in solutions space. These points are noted non-dominated solutions.

This part has been detailed in our work [20] [21].

## C. Calculation of lower bounds

Since we have transformed our multi-objective problem (MOP) in a mono-objective problem (PMOλ), as shown in Equation 1, which is to combine the different cost functions $f_i$ of the problem into a single objective function $F$, usually a linear [27].

The problem now, is to determine the various constants $c_i$ that are weights and scaling coefficients. The constants $c_i$ are usually initialized to $\frac{1}{f_i(x^*)}$ where $f_i(x^*)$ is the optimal solution associated to the objective function $f_i$.

Because we haven't information on the optimal solutions associated with different cost functions $f_i$ for our problem we are facing the calculation of lower bounds to determine the scaling constants $c_i$. To do this, we used the relaxation of various constraints.

All computation done, we obtain $f_{1b}$ and $f_{2b}$ successively representing a lower bound of the total travel cost and a lower bound of the total tardiness time. [26]

$$\begin{cases} \dfrac{c_1}{f_{1b}} \\ f_{1b} = \sum_{k \in K} C_k d_{min\,k.p.c} \end{cases} \quad (15)$$

$$\begin{cases} c_2 = \dfrac{1}{f_{2b}} \\ f_{2b} = \sum_{k \in K} \sum_{i \in N} \sum_{j \in N} max(0, D_i - l_i) X_{ijk} \quad (16) \\ s.c. \; f_{2b} \neq 0 \end{cases}$$

## D. Dynamic study

The dynamic m-pdptw is an extension of the m-pdptw where besides the temporal, precedence and capacity constraints, which imposes this last one, the dynamic m-pdptw distinguishes itself by the appearance of dynamic and urgent requests having planned the routing associated with each vehicle.

The planner will then have to satisfy the new requests besides the former while optimizing the chosen criteria and by respecting the imposed constraints.

We will present our approach of dynamic adaptation, by using two different methods for the insertion of new requests. The first method (Fig 2) consists in inserting the dynamic nodes from their appearances on the routing vehicle which minimizes the chosen criteria, by satisfying this last one. By keeping the same list of remaining nodes to visit, after insertion of dynamic requests.

The second method (Fig 3) is based on the adaptation of the genetic operators and the correction procedures of precedence and capacity initially conceived for the static case to serve in the dynamic case. While ensuring that vehicles must not, under any circumstances, visit a node already served.

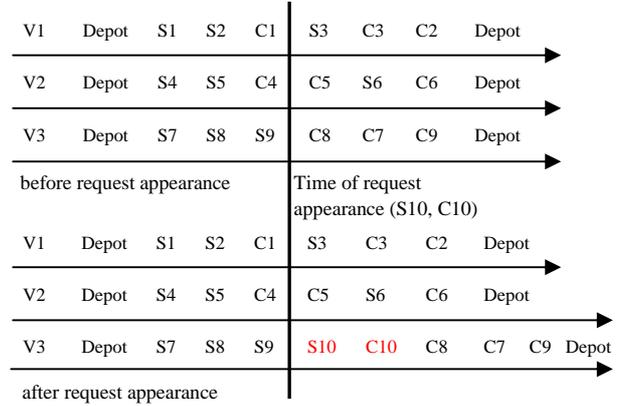

Figure 2: First method of insertion from dynamic request

Noting that the nodes which precede the appearance time of the dynamic request remain motionless.

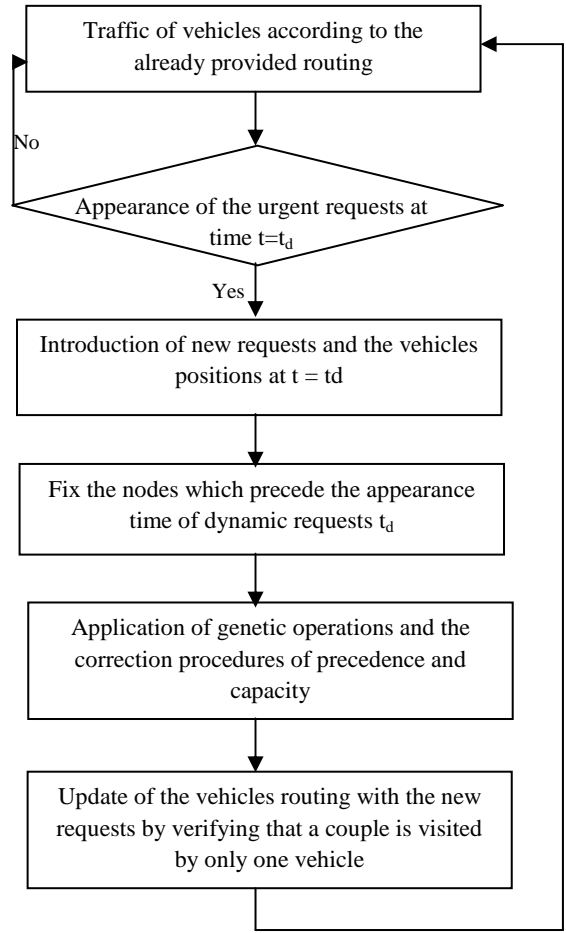

Figure 3: Second method of insertion from dynamic request

We present in Fig 4 our approach algorithm for the dynamic m-PDPTW.

---

**Begin**
**Step 1**: Create the initial population, (size n).
**Step 2**: Fill the intermediate population $P_{node}$ (size 2n) with individuals' crossover, mutation or copy.
**Step 3**: Correction procedure of Precedence and capacity.
**Step 4**: Create the 2$^{nd}$ intermediate population $P_{node/vehicle}$ (size 2n * 2n) representing the routing of each vehicle.
*If Appearance of the urgent requests at time $t=t_d$*
**Step 5**: Fix the nodes which precede the appearance time of dynamic requests $t_d$
*Where (the end criterion is not satisfied) do*
**Step 5.1**: Insert the dynamic nodes in each individual from the population $P_{node/vehicle}$
**Step 6.1**: Sort of population by the minimum value of fitness (Total tardiness time / Total travel cost)
**Step 7.1**: Copy non-dominated solutions for the first method of insertion from dynamic request
**End**
*Where (the end criterion is not satisfied) do*
**Step 5.2**: Insert the dynamic nodes in each individual from the population $P_{node/vehicle}$
**Step 6.2**: Create the 3$^{th}$ intermediate population $P_{node/vehicle\ dynamic}$ (size 2n*2n * 2n) with individuals' crossover, mutation or copy, for $t>t_d$
**Step 7.2**: Correction procedure of precedence and capacity and verify that a couple is visited by only one vehicle.
**Step 8.2**: Sort of population $P_{node/vehicle\ dynamic}$ by the minimum value of fitness (Total tardiness time / Total travel cost)
**Step 9.2**: Copy non-dominated solutions for the second method of insertion from dynamic request
**End**
**Else**: Traffic of vehicles according to the already provided routing
**End**

---

Figure 4: Approach algorithm for the dynamic m-PDPTW

### E. Computational results

To test our approach, we use benchmark problem instances generated by Li and Lim [17] from Solomon's ones [28]

Corresponding to Solomon's classification of C1, C2, R1, R2, RC1 and RC2, their data sets were also generated in six classes: LC1, LC2, LR1, LR2, LRC1 and LRC2. The LC problems are clustered whereas in the LR problems, providers and customers are randomly generated. Therefore in the LRC problems the providers and the customers are partially clustered and partially randomly distributed. While LC1, LR1 and LRC1 problems have a short scheduling horizon, LC2, LR2 and LRC2 have longer scheduling one. [29]

Each problem includes several classes. All these classes have approximately 100 suppliers and customers. They also contain the time windows, capacity, the quantities required for each vertex and the coupling constraints (supplier, customer).

In our work we will study the 8 classes that group the problem LRC1 noted successively LRC101 until LRC108.

Table I and II show the results of our simulation, by inserting it every time the couple of the dynamic request. Of course, for every given solution, we note the corresponding routing, crossed by each vehicle.

TABLE I: Results for the LRC1 problem with the First method of insertion from dynamic request

| LRC1 | $N_{sol}$ | $N_k$ | $f_1$ | $f_2$ | $F(x)$ |
|---|---|---|---|---|---|
| LRC101 | 5 | 25 | 234467,71 | 63,95 | 53,66 |
|  |  |  | 216616,68 | 73 |  |
| LRC101 | 47 | 9 | 170770,53 | 1489 | 0,49 |
|  |  | 11 | 213386,84 | 211,4 |  |
| LRC102 | 3 | 25 | 229200,37 | 64,2 | 53,47 |
|  |  |  | 233209,56 | 63,71 |  |
| LRC102 | 37 | 8 | 183345,04 | 914,9 | 0,49 |
|  |  | 12 | 224682,73 | 203,7 |  |
| LRC103 | 2 | 25 | 220334,76 | 63,77 | 53,52 |
|  |  |  | 212968,59 | 64,77 |  |
| LRC103 | 38 | 8 | 167587,35 | 677 | 0,49 |
|  |  | 10 | 196233,34 | 121,5 |  |
| LRC104 | 3 | 25 | 221505,56 | 64,21 | 53,47 |
|  |  |  | 246890,9 | 63,71 |  |
| LRC104 | 32 | 11 | 176883,78 | 412,8 | 0,49 |
|  |  | 12 | 206176,92 | 66,75 |  |
| LRC105 | 4 | 25 | 216597,48 | 70,07 | 53,66 |
|  |  |  | 248596,34 | 63,95 |  |
| LRC105 | 39 | 9 | 177702,26 | 771,9 | 0,49 |
|  |  | 12 | 215801,04 | 83,11 |  |
| LRC106 | 2 | 25 | 222740,68 | 64,14 | 53,39 |
|  |  |  | 231305,82 | 63,61 |  |
| LRC106 | 38 | 9 | 176349,14 | 644,8 | 0,49 |
|  |  | 12 | 205582,17 | 146,9 |  |
| LRC107 | 5 | 25 | 234829,59 | 63,53 | 53,32 |
|  |  |  | 220063,68 | 64,31 |  |
| LRC107 | 41 | 11 | 172050,35 | 286,6 | 0,49 |
|  |  | 12 | 211497,79 | 86,76 |  |
| LRC108 | 4 | 25 | 236116,92 | 63,92 | 53,64 |
|  |  |  | 221455,42 | 64,8 |  |
| LRC108 | 39 | 11 | 170105,82 | 421,9 | 52,09 |
|  |  | 13 | 224198,4 | 63,5 |  |

TABLE I: Results for the LRC1 problem with the Second method of insertion from dynamic request

| LRC1 | $N_{sol}$ | $N_k$ | $f_1$ | $f_2$ | $F(x)$ |
|---|---|---|---|---|---|
| LRC101 | 4 | 25 | 210071 | 8,2 | 2,29 |
|  |  |  | 217638,09 | 0 |  |
| LRC101 | 45 | 9 | 167285,5 | 3989 | 0,49 |
|  |  | 12 | 212849,42 | 143 |  |
| LRC102 | 4 | 25 | 224857,14 | 0 | 2,29 |
| LRC102 | 43 | 8 | 183345,04 | 849,8 | 0,49 |
|  |  | 14 | 221230,15 | 89,72 |  |
| LRC103 | 5 | 25 | 209584,25 | 0 | 2,29 |
|  |  |  | 209112,54 | 12,87 |  |
| LRC103 | 42 | 6 | 166937,7 | 3892 | 0,49 |
|  |  | 10 | 202398,45 | 57,63 |  |
| LRC104 | 10 | 25 | 216110,7 | 103,5 | 2,29 |
|  |  |  | 230606,14 | 1,38 |  |
| LRC104 | 39 | 11 | 213761,31 | 9,08 | 0,49 |
|  |  |  | 176834,1 | 84,95 |  |
| LRC105 | 18 | 25 | 231586,07 | 1,38 | 2,29 |
|  |  |  | 211112,71 | 142,2 |  |
| LRC105 | 36 | 9 | 177105,29 | 792,6 | 0,49 |
|  |  | 12 | 212829,1 | 70,58 |  |
| LRC106 | 3 | 24 | 220793,62 | 0 | 2,29 |
| LRC106 | 35 | 9 | 174037,5 | 1259 | 0,49 |
|  |  | 11 | 188774,51 | 59,49 |  |
| LRC107 | 3 | 25 | 215355,6 | 0 | 2,29 |
| LRC107 | 75 | 11 | 171035,53 | 222,6 | 0,49 |
|  |  | 12 | 212836,6 | 6,01 |  |
| LRC108 | 3 | 25 | 211749,64 | 0 | 2,29 |
| LRC108 | 45 | 9 | 195396,81 | 0 | 0,49 |
|  |  | 11 | 167210,96 | 357,3 |  |

With:

$N_{sol}$ : represents the number of non-dominated solutions.

$N_k$ : represents the vehicles number used.

Tables 1 and 2 show, for each study case, the solution which minimizes the total travel cost and the one who minimizes the total tardiness time. Noting that these two last ones are sorted among the not-dominated solutions.

Adding as well as for each case, we use two different methods for computation the vehicles number used. We consider a vehicle number k ranging between 1 and 25.

We observe that our approach generates a multiple number of solutions minimizing the compromise between the total travel cost and the total tardiness time, that give flexibility of choice for the decision maker.

We also observe that we obtain a total tardiness equal to zero with a tolerable cost, by using the second method of insertion from dynamic request.

These results justify the genetic algorithms utility.

## V. CONCLUSION

In this paper, we have presented our approach to solve the dynamic m-PDPTW, based on Pareto dominance method, with use of genetic algorithm and lower bounds. We proposed a brief literature review on the VRP, 1-PDPTW and m-PDPTW and The mathematical formulation of our problem. Then, we have detailed the use aggregation method and lower bounds to determine a set of solutions, minimizing our objective functions by inserting the couple of the dynamic request. Simulation was presented in a last part by using benchmark's data.